\documentclass{svjour2}

\usepackage{graphicx}

\journalname{Int J Theor Phys}

\begin{document}

\title{The higher dimensional gravastars}
\titlerunning{The higher dimensional gravastars}
\author{F. Rahaman \and S. Chakraborty \and Saibal Ray \and A. A. Usmani \and S. Islam}
\authorrunning{Rahaman \and Chakraborty \and Ray \and Usmani \and Islam}

\institute{F. Rahaman \at Department of Mathematics, Jadavpur
University, Kolkata 700032, India \email{rahaman@iucaa.ernet.in}
\and S. Chakraborty \at Department of Mathematics, Jadavpur
University, Kolkata 700032, India
\email{schakraborty@math.jdvu.ac.in} \and Saibal Ray \at
Department of Physics, Government College of Engineering \&
Ceramic Technology, Kolkata 700010, West Bengal, India
\email{saibal@iucaa.ernet.in} \and A.A. Usmani \at Department of
Physics, Aligarh Muslim University, Aligarh 202002, Uttar Pradesh,
India \email{anisul@iucaa.ernet.in} \and S. Islam \at Department
of Mathematics, Jadavpur University, Kolkata 700032, West Bengal,
India \email{sofiqul001@yahoo.co.in}}

\date{Received: date / Accepted: date}

\maketitle

\begin{abstract} A new model of {\it gravastar} is
obtained in $D$-dimensional Einstein gravity. This class of
solutions includes the gravastar as an alternative to
$D$-dimensional versions of the Schwarzschild-Tangherlini black
hole. The configuration of this new gravastar consists of three
different regions with different equations of state: [I] Interior:
$0 \leq r < r_1$, $ \rho = -p$; [II] Shell: $r_1 \leq r < r_2$, $
\rho = p$; [III] Exterior: $r_2 < r $, $ \rho = p =0$. The outer
region of this gravastar corresponds to a higher dimensional
Schwarzschild-Tangherlini black hole.

\end{abstract}

\section{Introduction}
Our recent model of a charge free gravastar in $(2+1)$-dimensional
anti-de Sitter spacetime~\cite{Rahaman2012a} and its subsequent
generalization to charged gravastars with electrovacuum exterior
\cite{Rahaman2012b} express perfectly the profound success of our
efforts in constructing a non-singular gravastar as an alternative
to  $(2+1)$ black holes. These efforts were inspired by an earlier
attempt by us~\cite{Usmani2011}, wherein we had constructed a
compact astrophysical charged object, a $(3+1)$-dimensional
charged gravastar, as an alternative to charged black holes.
However, present solution for the proposed astrophysical object of
higher dimensional gravastars is found to be singular at its
origin, which is a point of worry. Thus, there is a pertinent need
to understand the subject from the very basics of cleaner
$(2+1)$-dimensional gravity and develop the subject of these
gravitational vacuum stars starting from $(2+1)$-dimensions to
higher dimensions.

In connection to de Sitter spacetime and black holes a series of
works are available in the
literature~\cite{Dymnikova1992,Dymnikova1996,Dymnikova2000,Dymnikova2002,Dymnikova2007,Dymnikova2010,Dymnikova2011,Dymnikova2013}.
These investigations are interesting in the sense that the authors
have analyzed the globally regular solution of the Einstein
equations describing a black hole whose singularity is replaced by
the de Sitter core.

However, in our present study we extend the proposition of charge
free gravastars of Mazur and Mottola \cite{Mazur2001,Mazur2004} to
a charged compact object. While doing so, we invoke the very idea
of {\it electromagnetic mass} (EMM) which suggests that interior
de Sitter vacuum of a charged gravastar generates gravitational
mass \cite{Lorentz1904,Wheeler1962,Feynman1964,Wilczek1999}. This
provides a stable configuration by balancing the repulsive
pressure arising from charge with its alternative gravity to avert
a singularity.

It is a common trend to believe that the $4$-dimensional present
spacetime structure is the self-compactified form of manifold with
multidimensions. Therefore, cosmic string as well as superstring
theories and hence M-theory which reproduce higher dimensional
general relativity at low energy, argued that theories of
unification tend to require extra spatial dimensions to be
consistent with the physically viable models
\cite{Schwarz1985,Weinberg1986,Duff1995,Polchinski2007,Hellerman2007,Aharony2007}.
The classical analogue of the effective String Theory is the low
energy effective action containing squares and higher powers of
curvature terms. Also, similar higher derivative gravitational
terms appear in the renormalization of quantum field theory in
curved space background. Further, it is shown that some features
of higher dimensional black holes differ significantly from four
dimensional black holes as higher dimensions allow for a much
richer landscape of black hole solutions that do not have
$4$-dimensional counterparts \cite{Emparan2008}. It draws more
interest due to (1) a conceivable possibility of the production of
higher dimensional black holes in future colliders in the scenario
of large extra dimensions and TeV-scale gravity
\cite{Cavaglia2003,Kanti2004}, and  (2) The AdS/CFT correspondence
which relates the possibility of a $D$-dimensional black hole with
those of a quantum field theory in $(D-1)$-dimensions
\cite{Aharony2000}.

In fact, the study of higher dimensional black holes have gained
momentum in the first decade of this millennium. As in the present
paper we are considering gravastar as an alternative to black
holes so it is reasonable to adopt higher dimensional gravastar
due to importance of higher dimensional black holes. Therefore, we
present our study of higher dimensional gravastars proposed as an
alternative to higher dimensional Schwarzschild-Tangherlini black
holes \cite{Tangherlini1963}. We develop mathematical framework
for these gravastars and obtain solutions for its three separable
regions; the interior, the shell and the exterior. We then study
proper length and energy, entropy and junction conditions in
detail. The results and discussions have been presented in every
section under various headings and subheadings. At the end, we
conclude our findings.

\section{Interior space-time}
Since we are exploring for higher dimensional gravastar, we have
assumed a $D$-dimensional spacetime with the structure $R^1 X S^1
X S^d (d = D - 2)$, where $S^1$ is the range of the radial
coordinate $r$ and $R^1$ is the time axis. For this purpose, let
us consider a static spherically symmetric metric in $D = d + 2$
dimension as
\begin{equation}
ds^2 = -e^{\nu}dt^2 + e^{\lambda}dr^2+r^2 d\Omega_d ^2.\label{eq1}
\end{equation}

The notation, $ d\Omega_d ^2$ is a linear element on a
$d$-dimensional unit sphere, parametrized by the angles $\phi_1,
\phi_2,......,\phi_d:$ \\

$d\Omega_d ^2= d\phi_d^2 + \sin_2 \phi_d [d\phi_{d-1}^2 + \sin_2
\phi_{d-1}\{d\phi_{d-2}^2 + .........+ \sin_2 \phi_3(d\phi_2^2 +
\sin_2 \phi_2  d\phi_1^2).......\}] $. \\

The Hilbert action coupled to matter is given by
\begin{equation}
I = \int d^D x \sqrt{-g } \left( \frac{R_D}{16 \pi G_D} + L_{m}
\right),\label{eq2}
\end{equation}
where $R_D$ is the curvature scalar in $D$-dimensional spacetime,
$G_D$ denotes the $D$-dimensional Newton constant and $L_{m}$ is
the Lagrangian for matter distribution. We obtain the following
Einstein equation by varying the above action with respect to the
metric as
\begin{equation}
G^D_{ab}   = - 8 \pi G_D T_{ab},\label{eq3}
\end{equation}
where $G^D_{ab}$ denotes the Einstein's tensor in $D$-dimensional
spacetime.

The interior of the star is assumed to be perfect fluid type and
can be given by
\begin{equation}
T_{ij} = (\rho + p ) u_i u_j + p  g_{ij}, \label{eq4}
\end{equation}
where, $\rho$ represents the energy density, $p$ is  the isotropic
pressure, and  $u^{i}$ is the $D$-velocity of the fluid. The
Einstein field equations for the metric (\ref{eq1}), together with
the energy-momentum tensor given in Eq.~(\ref{eq2}), yield
\begin{equation}
-e^{-\lambda} \left[\frac{d(d-1)}{2r^2} - \frac{d \lambda'}{2r}
\right] + \frac{d(d-1)}{2r^2} = 8\pi G_D~ \rho, \label{eq5}
\end{equation}

\begin{equation}
e^{-\lambda} \left[\frac{d(d-1)}{2r^2} + \frac{d \nu'}{2r} \right]
- \frac{d(d-1)}{2r^2} = 8\pi G_D ~p, \label{eq6}
\end{equation}

\begin{eqnarray}
\frac{e^{-\lambda}}{2} \left[ \nu'' -\frac{\lambda'\nu'}{2}
+\frac{ {\nu'}^2}{2} -\frac{(d-1)(\lambda'-\nu')}{r} +
\frac{(d-1)(d-2) }{r^2} \right] \nonumber \\ - \frac{(d-1)(d-2)
}{2r^2} = 8\pi G_D~p,\label{eq7}~~~~~~~~~~~~
\end{eqnarray}

where a `$\prime$' denotes differentiation with respect to the
radial parameter $r$. Here we have assumed $c = 1$ in geometrical
unit. Conservation equation in $D$-dimensions implies
\begin{equation}
\frac{1}{2} \left(\rho + p\right)\nu' + p' =0. \label{eq8}
\end{equation}

Following Mazur-Mottola \cite{Mazur2001}, we assume the Equation
of State (EOS) for the interior region in the form
\begin{equation} p=-\rho. \label{eq9}\end{equation}

Using this EOS, one gets from Eq. (\ref{eq8})
\begin{equation}
\rho = constant =\rho_c,~~~(say). \label{eq10}
\end{equation}

We write this constant as, $\rho_c = d(d+1) \Lambda/16 \pi G_D$,
where $2 \Lambda/d(d+1)$ is the $D$-dimensional cosmological
constant. This means that in the interior we are essentially
considering the Cosmological Constant i.e. vacuum energy density
of Einstein \cite{Einstein1919,Zeldovich1972}.

Therefore, pressure may be expressed as follows
\begin{equation}
p= -\rho_c.\label{eq11}
\end{equation}

Using Eq. (\ref{eq9}) one gets the solutions of $\lambda$ from the
field Eq. (\ref{eq5}) as given below

\begin{equation}
 e^{-\lambda} = 1- \frac{16 \pi G_D \rho_c}{d(d+1)}r^2 +E r^{1-d},\label{eq12}
\end{equation}
where $E$ is an integration constant. Since $d>2$ and the solution
is regular at $r=0$, so we demand $E=0$.

Using Eq. (\ref{eq9}) one may obtain from Eqs. (\ref{eq5}) and
(\ref{eq6}), the following relation
\begin{equation}
  \ln C =\lambda + \nu,\label{eq13}
\end{equation}
where $\ln C$ is an integration constant. Thus  we have the
following interior solutions
\begin{equation}
 Ce^{-\lambda} =e^\nu = C( 1- \Lambda r^2).\label{eq14}
\end{equation}
We then calculate the active gravitational mass $M(r)$ in higher
dimensions, which is found to be
\begin{equation}
M(r) = \int_0^{{r_1=R}}~ \left[  \frac{2 \pi^{\frac{d+1}{2}}
}{\Gamma \left(\frac{d+1}{2}\right)}\right]r^d \rho dr =  \left[
\frac{2 \pi^{\frac{d+1}{2}} }{(d+1)\Gamma
\left(\frac{d+1}{2}\right)}\right] \rho_c ~R^{d+1}.\label{eq15}
\end{equation}

This is the usual gravitating mass for a $d$-dimensional sphere of
radius $R$ and energy density $\rho_c$. The space-time metric thus
obtained turns out to be free from any central singularity.

\section{Exterior space-time}
The exterior region  defined as ($p=\rho=0$) in higher dimensions
is nothing but a generalization of Schwarzschild solution, which
as obtained by Tangherlini \cite{Tangherlini1963} reads as

\begin{equation}
ds ^2 = - \left(1 - \frac{\mu}{ r^{d-1}}\right)  dt^2 + \left(1 -
\frac{\mu}{ r^{d-1}}\right)^{-1} dr^2 + d \Omega_d ^2.\label{eq16}
\end{equation}
Here $\mu=16\pi G_D M/\Omega_d$ is the constant of integration
with $M$, the  mass of the black hole and ${\Omega}_d$, the area
of a unit $d$-sphere as ${\Omega}_d = 2
\pi^{(\frac{d+1}{2})}/\Gamma(\frac{d+1}{2})$.

\section{Shell}
It is assumed that thin shell contains ultra-relativistic fluid of
soft quanta which obeys the EOS
\begin{equation} p =\rho. \label{eq17}\end{equation}

This represents stiff fluid model of Zel'dovich type in connection
to cold baryonic universe \cite{Zeldovich1972}.

It is difficult to obtain a general solution of the field
equations in the non-vacuum region, i.e. within the shell. We try
to find an analytic solution  within the thin shell limit,
 $0<e^{-\lambda}\equiv h <<1$.  As an advantage of it, we may set $h$ to
be zero to the leading order. Under this approximation, the field
Eqs.  (\ref{eq5}) - (\ref{eq7}), with the above EOS,  may be
recast in the following form
\begin{equation}
 \frac{  h'}{2 r}  = \frac{(d-1)}{r^2},\label{eq18}
\end{equation}

\begin{equation}
\frac{\nu'  h'}{4} +\frac{(d-1)h'}{2r}= -
\frac{(d-1)}{r^2}\label{eq19}.
\end{equation}

Integration of Eq. (\ref{eq18}) immediately yields
\begin{equation}
h = E+ 2 (d-1) \ln r,\label{eq20}
\end{equation}
where $E$ is an integration constant. The range of $r$ lies within
the thickness of the shell [$r_1=R, r_2=R+\epsilon$]. We, under
the condition  $\epsilon <<1$, get $E<<1$ as $h<<1$.

The other metric coefficient, $\nu$, can be found as
\begin{equation}
e^{\nu} =  \left(\frac{r}{r_0 }   \right)^{-2d},\label{eq21}
\end{equation}
where $r_0$ is an integration constant.

Also, from the conservation equation and using the same EOS as
above, one may obtain
\begin{equation}
p = \rho =  \rho_0 e^{-\nu} =\rho_0 \left(\frac{r}{r_0 }
\right)^{2d},\label{eq22}
\end{equation}
$\rho_0$ being an integration constant. As $\rho \propto r^{2d} $,
so the ultra relativistic matter in the shell ($r_1 \leq r < r_2$)
is more dense at the outer boundary ($r_2 < r $) than in the inner
boundary ($0 \leq r < r_1$).

\begin{figure}
\centering
\includegraphics[scale=.3]{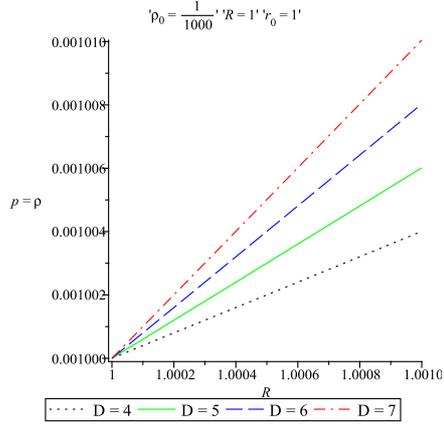}
\caption{The variation of pressure and density of the ultra
relativistic matter in the shell against $r$ for different
dimensions}\hfill
\end{figure}

\section{Proper length and Energy}
We consider matter shell is situated at the surface $ r = R $,
describing the phase boundary of region I. The thickness of the
shell ($\epsilon << 1$) is assumed to be very small. Thus the
region III joins at the surface $ r = R+\epsilon $.

Now, we calculate the proper thickness between two interfaces i.e.
of the shell as
\begin{equation}
\ell = \int _{R}^{R+\epsilon} \sqrt{e^{\lambda} } dr = \int
_{R}^{R+\epsilon} \frac{dr}{[E+ 2 (d-1) \ln
r]^{\frac{1}{2}}}.\label{eq23}
\end{equation}

By solving the above equation, one gets
\begin{equation}
\ell = \left[ a \frac{\sqrt{\pi} ~erf [\sqrt{-a
}\sqrt{(E+\frac{1}{a} \ln r)}]}{e^{aE}\sqrt{
-a}}\right]_R^{R+\epsilon},\label{eq24}
 \end{equation}
where $a =\frac{1}{2(d-1)}$.

\begin{figure}
\centering
\includegraphics[scale=.3]{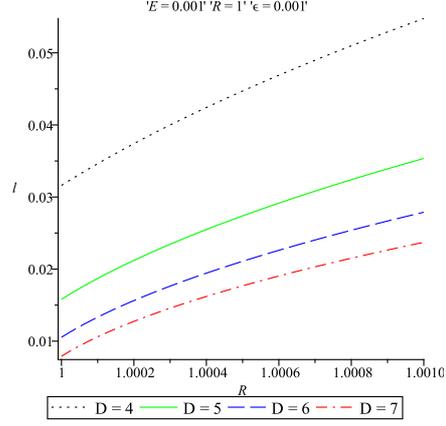}
\caption{The variation of proper length within the shell against
$r$ for different dimensions}
\end{figure}

It will be interesting to calculate the energy $\widetilde{E}$
within the shell, which we find out as

$\widetilde{E}  =   \int _{R}^{R+\epsilon}  \left[  \frac{2
\pi^{\frac{d+1}{2}} }{\Gamma \left(\frac{d+1}{2}\right)}\right]r^d
\rho dr =  \int _{R}^{R+\epsilon}  \left[  \frac{2
\pi^{\frac{d+1}{2}} }{\Gamma \left(\frac{d+1}{2}\right)}\right]r^d
\rho_0 \left(\frac{r}{r_0 } \right)^{2d} dr\\
\\
~~~~~~~~~~~~~~~~~~~~~~ = \left[  \frac{2 \pi^{\frac{d+1}{2}}
}{\Gamma \left(\frac{d+1}{2}\right)}\right]
 \left[\frac{\rho_0}{(3d+1)r_0^{2d}}\right]
 \left[(R+\epsilon)^{3d+1}-R^{3d+1}\right]$.\label{eq25}
\begin{equation} \end{equation}

However, one may write the energy $\widetilde{E}$ within the shell
up to first order in $\epsilon$ as
\begin{equation} \widetilde{E} \approx \left[  \frac{2 \pi^{\frac{d+1}{2}}
}{\Gamma \left(\frac{d+1}{2}\right)}\right] \rho_0 \left(
\frac{R}{r_0}\right)^{2d} R \epsilon. \label{eq26}\end{equation}

We observe that the energy within the shell is  not only
proportional to $\epsilon$ in first order of thickness but also
depends on dimension $d$ of the spacetime.

\section{Entropy}
We calculate the entropy following Mazur and Mottola prescription
\cite{Mazur2001} as
\begin{equation}
 S =  \int _{R}^{R+\epsilon} \left[  \frac{2
\pi^{\frac{d+1}{2}} }{\Gamma \left(\frac{d+1}{2}\right)}\right]r^d
s(r)   \sqrt{e^{\lambda}}dr.\label{eq27}
\end{equation}
Here, $s(r)$ stands for the entropy density of the local
temperature $T(r)$, which may be written as
\begin{equation}
s(r) =  \frac{\alpha^2k_B^2T(r)}{4\pi\hbar^2 } =
\alpha\left(\frac{k_B}{\hbar}\right)\sqrt{\frac{p}{2 \pi
}},\label{eq28}
\end{equation}
where $\alpha^2$ is a dimensionless constant.

Thus the entropy of the fluid within the shell could be found as
\begin{eqnarray}
S =   \int _{R}^{R+\epsilon}\left[  \frac{2 \pi^{\frac{d+1}{2}}
}{\Gamma \left(\frac{d+1}{2}\right)}\right]
 \sqrt{\frac{\alpha^2 \rho_0}{2 \pi
r_0^{2d}}}\left(\frac{k_B}{\hbar}\right) \frac{r^{2d} dr}{[E+ 2
(d-1) \ln r]^{\frac{1}{2}}}.\label{eq29}
\end{eqnarray}

Solving the above equation, one gets
\begin{equation}
S = \left[ \left(  \frac{2 \pi^{\frac{d+1}{2}} }{\Gamma
\left(\frac{d+1}{2}\right)}\right)
 \sqrt{\frac{\alpha^2 \rho_0}{2 \pi
r_0^{2d}}}\left(\frac{k_B}{\hbar}\right)\frac{1}{2(d-1)}
\frac{\sqrt{\pi} ~erf [\sqrt{-b }\sqrt{(E+2(d-1) \ln
r)}]}{e^{bE}\sqrt{ -b}}\right]_R^{R+\epsilon},\label{eq30}
\end{equation}
where $b =\frac{2d+1}{2(d-1)}$.

\begin{figure}
\centering
\includegraphics[scale=.3]{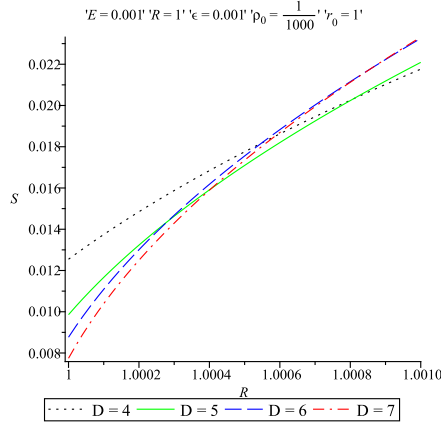}
\caption{The variation of Entropy within the shell against $r$ for
different dimensions}
\end{figure}

\section{Junction Condition}
 The gravastar configuration contains three regions in which
 interior region I is connected with exterior region II at the
 junction interface i.e. at the shell. This makes a geodecically
 complete manifold with a matter shell at the surface $r=R$. Thus
 a single manifold characterizes the gravastar configuration.
 According to fundamental junction condition there has to be a
 smooth matching between the regions I and III of the gravastar.
 However, though the metric coefficients are continuous at the
 junction surface ($S$) their derivatives may not be continuous there.
 Thus affine connections may be discontinuous at the boundary
 surface, in other words, the second fundamental forms \cite{Israel1966,Rahaman2006,Rahaman2009,Usmani2010,Rahaman2010,Dias2010,Rahaman2011}

 \begin{equation}K_{ij}^\pm = - n_\nu^\pm\ \left[\frac{\partial^2x^\nu}
{\partial \xi^i\partial \xi^j } +
 \Gamma_{\alpha\beta}^\nu \frac{\partial x^\alpha}{\partial \xi^i}
 \frac{\partial x^\beta}{\partial \xi^j }\right]_{ |_S},\label{eq31}
 \end{equation}
where, $ n_\nu^\pm\ $ are the unit normals to $S$ and can be
written as
\begin{equation} n_\nu^\pm =  \pm   \left| g^{\alpha\beta}\frac{\partial f}{\partial x^\alpha}
 \frac{\partial f}{\partial x^\beta} \right|^{-\frac{1}{2}} \frac{\partial f}{\partial
 x^\nu}~~~~~~~~~~with~~~~~~~~~ n^\mu n_\mu ~ = ~1,\label{eq32}
 \end{equation}
which are associated with the two sides of the shell are
discontinuous.

In Eq. (\ref{eq29}), $\xi^i$ are the intrinsic coordinates on the
shell and $f(x^\alpha(\xi^i)) =0$ is the parametric equation of
the shell $S$. Here, $-$ and $+$ mention interior and exterior
regions.

These discontinuity of the second fundamental forms,
\begin{equation}\kappa _{ij} = K^+_{ij}-K^-_{ij},\label{eq33}
 \end{equation}
produce intrinsic stress energy tensor within the shell. Using
Lanczos equations
\cite{Lanczos1924,Sen1924,Darmois1927,Israel1966,Perry1992,Musgrave1996},
one can write the surface intrinsic energy momentum tensors,
$S_i^j = diag (-\sigma, -v, -v, .....,-v) $ where

\begin{equation}\sigma = -\frac{1}{8\pi G_D}  \kappa _\tau^\tau,\end{equation}
is the surface energy density and
\begin{equation}-v = \frac{1}{8\pi G_D}  \kappa _{\phi_A}^{\phi_A},\label{eq34}
\end{equation}

is the surface tension.

For our gravastar configuration, we calculate
\begin{equation} \sigma =  -\frac{d}{8\pi G_DR}  \left[ \sqrt{
1-\frac{\mu}{R^{d-1}}} - \sqrt{ 1-  \Lambda
R^2}\right],\label{eq35}
\end{equation}

\begin{equation} v =  - \frac{1}{8\pi G_D}  \left[ \frac{\frac{\mu (d-1)}{2
R^{d-1}} + (d-1) (1-\frac{\mu}{R^{d-1}})} {\sqrt{
1-\frac{\mu}{R^{d-1}}}}-\frac{-\Lambda R^2 +(d-1)(1-\Lambda R^2) }
{\sqrt{1-\Lambda R^2}}\right].\label{eq36}
\end{equation}

We see that the energy density as well as surface tension of the
junction shell are negative. This means we have a thin shell of
matter content with negative energy density. It is to be noted
that the discontinuity of the affine connections at the region II
i.e. in the shell provides the above matter confined within the
shell. Such a stress-energy tensor is not ruled out from the
consideration of Casimir effect between compact objects at
arbitrary separations \cite{Emig2007}. The above negative surface
tension also indicates that there is a surface pressure as opposed
to surface tension. Thus, in principle, the shell of our gravastar
configuration consists of a combination of two types of matter
distributions, namely, the ultra-relativistic fluid obeying
$p=\rho$ and matter components due to discontinuity of second
fundamental form of the junction interface, that are given in Eqs.
(\ref{eq35}) and (\ref{eq36}). We demand that these two fluids are
non-interacting and characterize the shell of the gravastar i.e.
non-vacuum region II.

\section{Concluding remarks}
In the present work we generalize the concept of gravastar, {\it a
gravitational vacuum star}, in the spacetime of $4$-dimensional to
$D$–dimensional Einstein gravity of the Schwarzschild-Tangherlini
category black hole. To do so, {\it firstly}, we have considered
three different regions with different EOS such as [I] $0 \leq r <
r_1$, $ \rho = -p$ (Interior), [II] $r_1 \leq r < r_2$, $ \rho =
p$ (Shell) and [III] $r_2 < r $, $ \rho = p =0$ (Exterior). {\it
Secondly}, the conjecture of electromagnetic mass (EMM) has been
invoked due to the presence of charge. Originally Lorentz
\cite{Lorentz1904} proposed model for extended electron and
conjectured that ``there is no other, no `true' or `material'
mass,'' and thus provides only `electromagnetic masses of the
electron'. Wheeler \cite{Wheeler1962} and Wilczek
\cite{Wilczek1999} also argued that electron has a ``mass without
mass''. Feynman, Leighton and Sands \cite{Feynman1964} termed this
type of models as ``electromagnetic mass models''. Following the
idea of EMM, where all the physical parameters, including the
gravitational mass, are arising from the electromagnetic field
alone, have been extensively studied by several investigators
\cite{Florides1962,Cooperstock1978,Tiwari1984,Gautreau1985,Gron1985,Leon1987,Ray2004,Ray2006,Ray2008}
under the general relativistic framework where spacetime geometry
is assumed to be associated with the presence of charged particle
obeying Maxwell's equations of electromagnetic theory.

However, in connection to the {\it interior configuration I} of
EMM we would like to record that most of the above investigators
exploit an EOS with a repulsive pressure of the form $p = -\rho$
which is a very common feature in the context of the present
accelerating Universe and have been argued to be connected with
$\Lambda$-dark energy
\cite{Perlmutter1998,Riess1998,Ray2007,Usmani2008}. The EOS of
this type implies that the matter distribution under consideration
is in tension and hence the matter is known in the literature as a
`false vacuum' or `degenerate vacuum' or `$\rho$-vacuum'
\cite{Davies1984,Blome1984,Hogan1984,Kaiser1984}. This EOS was
first discussed by Gliner \cite{Gliner1966} in his study of the
algebraic properties of the energy-momentum tensor of ordinary
matter through the metric tensors. Later on it was revealed that
the gravitational effect of the zero-point energies of particles
and electromagnetic fields are real and measurable, as in the
Casimir Effect \cite{Casimir1948}.

Whereas in connection to the {\it shell configuration II} it is to
note that the stiff fluid model, which refers to a Zel'dovich
universe, have been employed by several authors for various
situations such as cold baryonic universe \cite{Zeldovich1972},
early hadron era \cite{Carr1975}, scalar field fluid
\cite{Madsen1992} and LRS Bianchi-I cosmological models
\cite{Chakrabarty2001}. There are also recent applications and
claims for stiff fluid EOS in the various astrophysical systems
like neutron star RX J1856-3754 \cite{Braje2002}, hyperon stars
\cite{Linares2004} and structure formation \cite{Buchert2001}.

As a final remark we would like to add here that our sole aim in
the present work was to find a classical analogue of the higher
dimensional gravastar as an alternative to black holes and it
seems that we are quite successful in our attempt.

\section*{Acknowledgments}
FR, SC, SR and AAU wish to thank the authorities of the
Inter-University Centre for Astronomy and Astrophysics, Pune,
India for providing the Visiting Associateship under which a part
of this work was carried out. FR is also thankful to UGC for
providing financial support. We all are grateful to the referee
for helpful suggestions which made us to upgrade the manuscript in
a substantial manner.\\

\end{document}